\pdfoutput=1
\documentclass[manuscript]{acmart}
\usepackage[utf8]{inputenc}
\usepackage{multirow}
\usepackage{amsmath}
\usepackage{cleveref}
\usepackage{newtxmath}

\DeclareMathOperator*{\argmin}{arg\,min}

\title{Are Quantum Computers Practical Yet? A Case for Feature Selection in Recommender Systems using Tensor Networks}
\date{April 2022}

\usepackage{bm}

\newcommand{\ICM}{ICM}
\newcommand{\IPM}{IPM}
\newcommand{\K}{IPM_{reward}}
\newcommand{\E}{IPM_{penalty}}
\newcommand{\FPM}{FPM}
\newcommand{\BQM}{BQM}

\newcommand{\fat}[1]{\ifmmode\bm{#1}\else\textbf{#1}\fi}

\newcommand{\set}[1]{\mathbb{#1}}

\newcommand{\vect}[1]{\fat{#1}}

\newcommand{\matr}[1]{#1}

\newcommand{\tens}[1]{\mathcal{#1}}

\newcommand{\func}[1]{\textsf{#1}}

\newcommand{\order}[1]{\mathcal{O}\left( #1 \right)}

\newcommand{\my}[0]{\matr{Y}}            

\newcommand{\tg}[0]{\tens{G}}            
\newcommand{\ty}[0]{\tens{Y}}            

\begin{document}

\author{Artyom Nikitin}
\email{artyom.nikitin@skoltech.ru}
\affiliation{%
  \institution{Skolkovo Institute of Science and Technology}
  \city{Moscow}
  \country{Russian Federation}
}

\author{Andrei Chertkov}
\email{andrei.chertkov@skoltech.ru}
\affiliation{%
  \institution{Skolkovo Institute of Science and Technology}
  \city{Moscow}
  \country{Russian Federation}
}

\author{Rafael Ballester-Ripoll}
\email{rballester@faculty.ie.edu}
\affiliation{%
  \institution{School of Science and Technology, IE University}
  \city{Madrid}
  \country{Spain}
}

\author{Ivan Oseledets}
\email{i.oseledets@skoltech.ru}
\affiliation{%
  \institution{Skolkovo Institute of Science and Technology}
  \city{Moscow}
  \country{Russian Federation}
}
\affiliation{%
  \institution{Artificial Intelligence Research Institute}
  \city{Moscow}
  \country{Russian Federation}
}

\author{Evgeny Frolov}
\email{evgeny.frolov@skoltech.ru}
\affiliation{%
  \institution{Skolkovo Institute of Science and Technology}
  \city{Moscow}
  \country{Russian Federation}
}

\begin{abstract}
Collaborative filtering models generally perform better than content-based filtering models and do not require careful feature engineering. However, in the cold-start scenario collaborative information may be scarce or even unavailable, whereas the content information may be abundant, but also noisy and expensive to acquire. Thus, selection of particular features that improve cold-start recommendations becomes an important and non-trivial task. In the recent approach by Nembrini et al., the feature selection is driven by the correlational compatibility between collaborative and content-based models. The problem is formulated as a Quadratic Unconstrained Binary Optimization (QUBO) which, due to its NP-hard complexity, is solved using Quantum Annealing on a quantum computer provided by D-Wave. Inspired by the reported results, we contend the idea that current quantum annealers are superior for this problem and instead focus on classical algorithms. In particular, we tackle QUBO via TTOpt, a recently proposed black-box optimizer based on tensor networks and multilinear algebra. We show the computational feasibility of this method for large problems with thousands of features, and empirically demonstrate that the solutions found are comparable to the ones obtained with D-Wave across all examined datasets.
\end{abstract}

\maketitle

\section{Introduction}

\subsection{Quantum Optimization for Recommender Systems}

Quantum computing is a rapidly emerging technology. Quantum labs at industry giants such as Microsoft, Google, and IBM are actively working on both hardware and algorithmic aspects towards practical quantum solutions. The key driver for their innovation efforts is the technology's promise for substantial computational gains across multiple domains: information retrieval, financial analysis, cryptography, and more. 

There are currently two main lines of research in quantum computing. \emph{Gated} quantum computers are Turing-complete and thus general-purpose, but they have not yet reached commercial viability due to outstanding engineering challenges. 
\emph{Quantum annealers}, on the other hand, are more mature. For example, Google in partnership with D-Wave Systems Inc. already provide the Advantage chip, a 5000-qubit annealer~\cite{WWGJDSM:22}, and plan to scale it to the million-qubit range within the next 10 years. The main attractive point of quantum annealers lies in their potential to tackle large optimization problems more efficiently than classical computers (i.e. the so-called \emph{quantum speed-up}). Indeed, many business problems can be modeled as combinatorial optimization tasks with exponentially large solution spaces.

Besides numerous applications in fields like operations research (including traffic routing or task scheduling~\cite{DSPW:21}), quantum computing has been applied to recommender systems in the last few years~\cite{KP:17, ADBL:20}. Recently, a new line of research~\cite{nembrini2021feature, FFC:21} proposes to leverage quantum annealing for feature and model selection in recommender systems; applications include overcoming the \emph{cold-start} problem and generating \emph{recommendation carousels}. These approaches make use of D-Wave's Advantage annealing chip. To this end, the authors first cast the problem at hand as a Quadratic Unconstrained Binary Optimization (QUBO), a problem readily supported by the Advantage. Since some selection problems considered by~\cite{nembrini2021feature} involve up to 8'000 features and do not fit into the chip, the authors resort in these cases to a classical-quantum hybrid scheme that first splits the problem into smaller pieces.



\subsection{Practical Challenges}

Like their gated counterparts, quantum annealers present ongoing engineering challenges. For example, D-Wave's chips require superconductivity at extremely low temperatures, which brings operational costs to the order of thousands of dollars per hour. From the algorithmic point of view, current quantum annealers suffer from two kinds of limitations:

\begin{itemize}
    \item Current architectures for QUBO impose sparsity constraints on the interaction matrix's coefficients. For instance, each qubit in the Advantage chip is only connected to 15 other qubits. This shortcoming can be largely circumvented by \emph{minor embedding}, a technique that increases the number of connections available to each logical qubit by mapping it to multiple physical qubits. However, this process increases the overall number of physical qubits required by the solver; in addition, embedding is itself an NP-hard problem. An alternative approach is to use hybrid classical-quantum solvers~\cite{UNM:17}.
    \item \emph{Tunneling} is a quantum phenomenon that allows the annealer configuration to easily jump around the solution space, potentially helping it avoid valleys and local minima in the optimization landscape. Despite this, and as is the case with classical simulated annealing (SA), the process can still lead to sub-optimal solutions after a finite amount of time, and it is best to collect multiple low-energy samples in the hope of obtaining a high-quality solution. D-Wave's 2000Q chip, Advantage's predecessor, was recently compared with classical SA for QUBO with inconclusive results~\cite{KOKT:21}.
\end{itemize}

\subsection{Proposed approach}

In this paper, we set out to assess the attractiveness of quantum annealing for feature selection by comparing it to a classical (in a sense of ``non-quantum'') black-box gradient-free optimization algorithm TTOpt (Tensor Train Optimizer)~\cite{sozykin2022ttopt} that combines the power of \emph{tensor train} (TT) decomposition~\cite{Oseledets:11} with \emph{cross-approximation} \cite{oseledets2010ttcross} and maximum volume submatrix construction techniques \cite{goreinov2010find}.
TTOpt demonstrates high performance for optimization of essentially multidimensional functions and arrays. It was successfully applied in~\cite{sozykin2022ttopt} for neural network-optimization problems and demonstrated superiority compared to evolutionary algorithms and Bayesian methods. Thus, potentially the TTOpt approach can be applied to complex QUBO problems with thousands and tens of thousands of features, and it means we can cover many realistic scenarios for feature selection task in recommender systems.
\section{Problem Description}\label{sec:problem}

In this work we follow the Collaborative-driven Quantum Feature Selection (CQFS) approach proposed by Nembrini et al. \cite{nembrini2021feature}. The main idea is to select features in such a way that the item similarity matrix $S^{CBF}$ of a content-based filtering model approximates the item similarity matrix $S^{CF}$ of a collaborative filtering model. They formulate the problem as a Quadratic Unconstrained Binary Optimization (QUBO), efficiently solvable with quantum computers, and propose the following pipeline:
\begin{enumerate}
    \item Select, build and optimize particular collaborative and content-based models. For these models compute item similarity matrices with positive entries $S^{CBF} \in \mathbb{R}^{I \times I}$ and $S^{CF} \in \mathbb{R}^{I \times I}$, where $I$ is the number of items.
    \item Construct an \emph{Item Penalty Matrix} $\IPM \in \mathbb{R}^{I \times I}$, which penalizes inconsistencies between similarity matrices and rewards consistencies.
    \item Build a \emph{Feature Penalty Matrix} $\FPM \in \mathbb{R}^{F \times F}$ from $\IPM$ to penalize features instead of items, where $F$ is the number of features to select from.
    \item As an only a particular number of features may be important to select – add a soft constrained $c$ penalizing deviations from the desired amount and obtain the $BQM$ matrix.
    \item Solve the QUBO problem by finding the minimum of $x^T BQM x$, where $x \in \{0, 1\}^F$ is a vector of selected features.
    \item Finally, build and optimize a content-based model using the selected features.
\end{enumerate}
More details behind the idea of the proposed approach can be found in the original paper \cite{nembrini2021feature}. Now we will describe the steps taken to formulate the QUBO problem. Assume that we have build both collaborative- and content-based models and acquired their item similarity matricies $S^{CF}$ and $S^{CBF}$, respectively. Then, we want the Item Penalty Matrix $IPM$ built in such a way that for any item with indices $i, j \in {1, ..., I}$:
\begin{itemize}
    \item $\IPM_{ij} < 0$: if $S_{ij}^{CBF} \not\approx 0$ and $S_{ij}^{CF} \not\approx 0$ – consistent and we reward it;
    \item $\IPM_{ij} > 0$: if $S_{ij}^{CBF} \not\approx 0$ and $S_{ij}^{CF} \approx 0$ – may be possible to remove features and reduce item correlations to make consistent, hence, penalize it;
    \item $\IPM_{ij} = 0$: if $S_{ij}^{CBF} \approx 0$ and $S_{ij}^{CF} \not\approx 0$ – can not remove features to make consistent as items do not have any in common; 
    \item $\IPM_{ij} = 0$: if $S_{ij}^{CBF} \approx 0$ and $S_{ij}^{CF} \approx 0$ – consistent.
\end{itemize}

IPM is created as a composition of two matrices to balance penalization of inconsistent and rewarding of consistent entries:
\begin{equation}\label{eq:ipm}
    \IPM = \K + \beta \cdot \E,
\end{equation}
where $\K \in \{0, -1\}^{I \times I}$ describes rewarding and $\E \in \{0, 1\}^{I \times I}$ refers to penalization. To transmute item penalization to feature penalization and construct a Feature Penalty Matrix $FPM$ – some connection between features and items must be established. To that end, the authors use an Item Content Matrix $\ICM \in \{0, 1\}^{I \times F}$ denoting features used by each item. Therefore, $FPM \in \mathbb{R}^{F \times F}$ can be written as
\begin{equation}\label{eq:fpm}
    \FPM = \ICM^T \cdot \IPM \cdot \ICM
\end{equation}
Since the $\IPM$ matrix can be very large and it depends on the parameter $\beta$, it is more practical to compute $\FPM$ as follows:
\begin{equation}\label{eq:fpm_split}
\begin{split}
    \FPM & = \ICM^T \cdot \IPM \cdot \ICM = \\
    & = \ICM^T \cdot \K \cdot \ICM + \beta \cdot \ICM^T \cdot \E \cdot \ICM =\\
    & = \FPM_{reward} + \beta \cdot \FPM_{penalty}
\end{split}
\end{equation}

Now we can write the QUBO problem, additionally incorporating the soft constraint for the percentage of the features we would like to select:
\begin{equation}\label{eq:qubo_constraint}
    \argmin_{x~\text{binary}}~x^T \FPM x + c, \quad
    \text{where} \,\,
    c = s (\mathbf{1}^Tx - pF)^2,
\end{equation}
$s$ is a scaling factor, $\mathbf{1}$ is a vector of ones, and $p$ is the fraction of the desired amount of features $F$. We can additionally simplify this formulation and get rid of the second term:

\begin{equation}
    \BQM = \FPM + s \mathbf{1} \mathbf{1}^T - 2spF\vmathbb{1},
\end{equation}
where $\vmathbb{1}$ is an identity matrix. Finally,

\begin{equation}\label{eq:qubo}
    \argmin_{x~\text{binary}}~x^T \FPM x + c = \argmin_{x~\text{binary}}~x^T \BQM x.
\end{equation}
Note that, as we will show in the results \cref{sec:results}, such a soft constraint on the number of features may not result in selecting the exact or even approximate amount of features desired.

While Nembrini et al.~\cite{nembrini2021feature} compared the performance of a quantum solver with that of classical algorithms for the QUBO problem, we propose to use a different classical optimizer that is inspired by tensor networks. The type of network we choose, the TT-decomposition, is known as a matrix product states (MPS) in the quantum community and has been recently benchmarked with quantum methods for another optimization problem~\cite{MKSFLLO:22}. Our approach is detailed in the following section.
\section{Proposed Approach}

{
TTOpt (Tensor Train Optimizer)~\cite{sozykin2022ttopt} is a promising method for gradient-free global optimization of multivariable functions and discrete multidimensional arrays.
TTOpt is inspired by three recent methods in modern computational mathematics: low-rank TT-decomposition~\cite{oseledets2011tensor}, maximum volume submatrix construction~\cite{goreinov2010find}, and the multidimensional cross approximation method in the TT-format (TT-CAM)~\cite{oseledets2010ttcross}.
In this section, we briefly describe the main idea of the TTOpt, focusing on specific features for our task of the QUBO problem optimization.
}

\subsection{Tensor networks and low rank approximations}

In the formalism of tensor networks, a tensor $\ty \in \set{R}^{N_1 \times N_2 \times \ldots \times N_d}$ (we will use upper case calligraphic letters for tensors below) is just a multidimensional array with a number of dimensions $d$ and mode sizes $N_1, N_2, \ldots N_d$. For $d = 2$ and $d = 1$ we will also call them matrices and vectors, respectively. The size of $\ty$ is exponential in the number of dimensions $d$, and the tensor cannot be evaluated or stored for sufficiently large $d$.
Fortunately, efficient approximations were developed to work with multidimensional tensors in recent years (see, e.g.,~\cite{cichocki2016tensor, cichocki2017tensor}).

\begin{figure}[t!]
    \centering
    \includegraphics[width=0.95\textwidth]{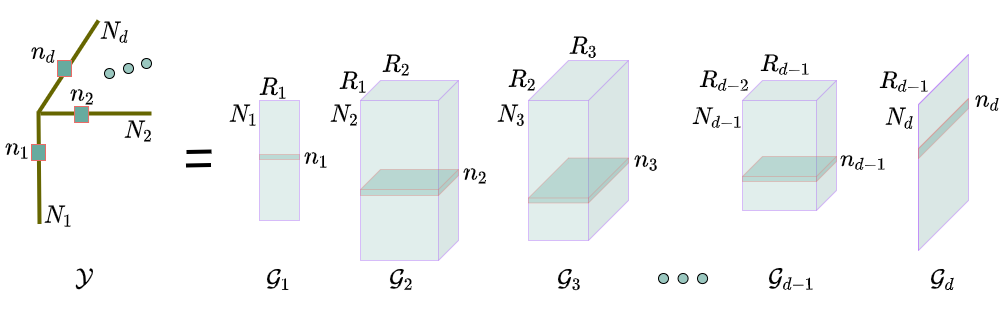}
    \caption{
        Schematic representation of the low-rank TT-decomposition.
        To compute an arbitrary element $\vect{n} = [n_1, n_2, \ldots, n_d]$ of the tensor $\ty \in \set{R}^{N_1 \times N_2 \times \ldots \times N_d}$, the convolution of the TT-cores $\tg_1, \tg_2, \ldots, \tg_d$ according to the formula~\eqref{eq:tt-repr-mtr} should be performed.
    }
    \label{fig:tt-element}
\end{figure}

A very promising low rank (low parametric) tensor format is provided by the TT-decomposition~\cite{oseledets2011tensor}.
A tensor $\ty \in \set{R}^{N_1 \times N_2 \times \ldots \times N_d}$ is said to be in the TT-format, if any its element can be represented by the following formula (see also illustration on Figure~\ref{fig:tt-element}):
\begin{equation}\label{eq:tt-repr-mtr}
\ty [n_1, n_2, \ldots, n_d]
=
\matr{G}_1(n_1)
\matr{G}_2(n_2)
\ldots
\matr{G}_d(n_d),
\quad
n_k = 1, 2, \ldots, N_k
\,\,
(k = 1, 2, \ldots, d),
\end{equation}
where $\matr{G}_k(n_k) = \tg_k [:, n_k, :]$ is an $R_{k-1} \times R_k$ matrix for each fixed $n_k$. The 3D tensors $\tg_k \in \set{R}^{R_{k-1} \times N_k \times R_k}$ are called TT-cores, while integers $R_{0}, R_{1}, \ldots, R_{d}$ (with the convention $R_{0} = R_{d} = 1$) are named TT-ranks.
Storing the TT-cores $\tg_1, \tg_2, \ldots, \tg_d$ requires less or equal than $d \hat{N} \hat{R}^2$ memory cells instead of $\hat{N}^d$ cells for the uncompressed tensor\footnote{
    We denote here and below the maximum size of the tensor mode as $\hat{N} = \max_{1 \leq k \leq d}{N_k}$ and the maximum TT-rank as $\hat{R} = \max_{1 \leq k < d}{R_k}$.
}.
Hence, the TT-decomposition is free from the curse of dimensionality if the TT-ranks are bounded.

The TT-decomposition can be computed\footnote{
    To construct a TT-approximation of a tensor $\widehat{\ty}$ means to choose TT-ranks $R_{1}, R_{2}, \ldots, R_{d-1}$ and 3D TT-cores $\tg_1, \tg_2, \ldots, \tg_d$ such that $\ty \approx \widehat{\ty}$.
} via standard linear algebra operations (such as SVD and QR)~\cite{oseledets2009breaking} from the given explicit full tensor.
However, for high-dimensional tensors this turns out to be too computationally complex; moreover, storing the original tensor in a full format is impossible in the first place if, say, $d \gg 10$.
To solve this problem, the TT-CAM approach was developed.

\subsection{Multidimensional cross approximation and the maximal volume principle}

The TT-CAM~\cite{oseledets2010ttcross, dolgov2020parallel} produces a TT-approximation $\ty$ (see formula~\eqref{eq:tt-repr-mtr} and Figure~\ref{fig:tt-element}) of the tensor $\widehat{\ty} \in \set{R}^{N_1 \times N_2 \times \ldots \times N_d}$, given implicitly as a function $\func{f}(n_1, n_2, \ldots, n_d)$ that returns the $(n_1, n_2, \ldots, n_d)$-th entry of $\widehat{\ty}$ for any given set of multi-indices.
The TT-CAM approach is based on the idea of applying a well-established CAM algorithm~\cite{goreinov2010find, CAIAFA2010557} to successive unfolding matrices\footnote{
    The $k$-th unfolding matrix $\my_k$ for the $d$-dimensional tensor $\ty \in \set{R}^{N_1 \times N_2 \times \cdots \times N_d}$ is the matrix $\my_k \in \set{R}^{N_1 \ldots N_k \times N_{k+1} \ldots N_d}$ with elements (for all possible indices):
    $
    \my_k[\,
        \overline{n_1, \ldots, n_k},
        \overline{n_{k+1}, \ldots, n_d}
    \,] \equiv \ty[n_1, n_2, \ldots, n_d]
    $.
} of an implicitly specified tensor.
Note that during the iterations of the TT-CAM, only a small, adaptively chosen, part of the original tensor is requested.

CAM for matrices (also called cross or skeleton decomposition) may be built iteratively by alternating directions method and a maximum volume (\func{maxvol}) algorithm~\cite{goreinov2010find}.
The maxvol algorithm finds $R$ rows in an arbitrary non-degenerate matrix $\matr{A} \in \set{R}^{N \times R}$ ($N > R$) which span a maximal-volume $R \times R$ submatrix $\hat{\matr{A}}$.
This submatrix $\hat{\matr{A}} \in \matr{A}$ has maximal value of the modulus of the determinant on the set of all nondegenerate square submatrices of size $R \times R$.
The algorithm starts from some initial approximation (it can be obtained, for example, from the LU decomposition), and then greedily rearranges rows of $\matr{A}$ to maximize submatrix volume.
Its computational complexity is $O(N R^2 + K N R)$, where $K$ is a number of iterations.

Within the framework of the CAM, successive maximal-volume submatrices for the rows and columns (``crosses'') of the given rectangular matrix are constructed until convergence. The rows and columns during iterations are selected in accordance with the selected indices that form the maximal-volume submatrix.
It was proved in~\cite{goreinov2010find} that if $\matr{\hat{A}}$ is an $R \times R$ submatrix of maximal volume (in selected rows and columns) of the matrix $\matr{A} \in \set{R}^{N_1 \times N_2}$, then the maximal (by modulus) element $\hat{a}_{max} \in \matr{\hat{A}}$ bounds the absolute maximal element $a_{max}$ in the full matrix $\matr{A}$:
\begin{equation}\label{eq:maxvol_maxelement}
    \hat{a}_{max} \geq \frac{a_{max}}{R^2}.
\end{equation}
Note that this bound is pessimistic, and in practice, the maximal-volume submatrix may contain the element which is very close to the optimal one.
Since choosing a submatrix with a large volume is more straightforward than finding the element with the largest absolute value and allows to take into account the internal correlations of tensor elements, this idea yields an efficient black-box discrete optimization method, as will be discussed below.

\subsection{Gradient-free discrete multidimensional optimization}

\begin{figure}[t!]
    \centering
    \includegraphics[width=\textwidth]{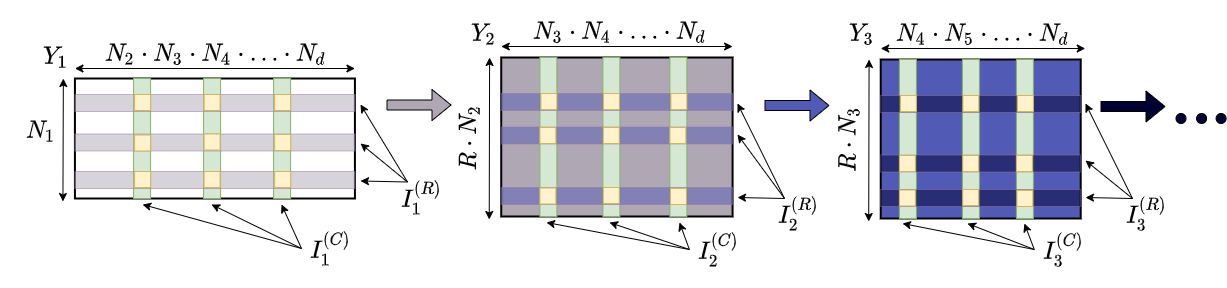}
    \caption{
        Conceptual scheme of TTOpt iterations.
        Algorithm starts from the first unfolding $\my_1$ of implicitly given tensor $\ty \in \set{R}^{N_1 \times N_2 \times \ldots \times N_d}$.
        According to a given set of multi-indices $I_1^{(C)}$, the corresponding columns are computed (shown in green; for compactness in this figure, we omit the transformation of column data, see Figure~\ref{fig:ttopt_scheme_ext} for details), then the rows $I_1^{(R)}$ (shown in purple) corresponding to the maximal-volume submatrix (shown in yellow) are determined, and then the submatrix formed by $I_1^{(R)}$ is converted into the following unfolding matrix $\my_2$.
        This process continues until the last mode $d$ of the tensor is reached, and then it is carried out in the opposite direction, i.e., from the last to the first mode.
        }
    \label{fig:ttopt_scheme}
\end{figure}

\begin{figure}[t!]
    \centering
    \includegraphics[width=\textwidth]{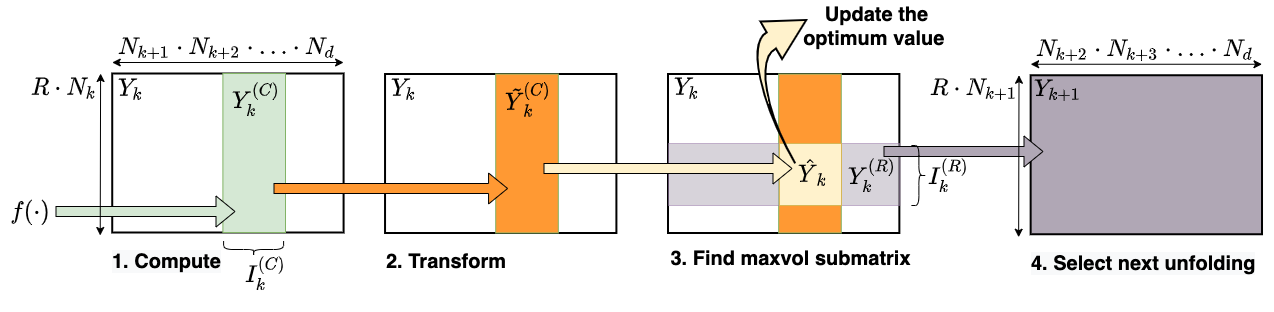}
    \caption{
        The sequence of operations performed by the TTOpt algorithm while forward stage of the sweep for the $k$-th unfolding matrix ($k = 1, 2, \ldots, d$; note that for the case $k = 1$, the number of rows in $\my_1$ will be $N_1$, not $R \cdot N_1$).
        For the simplicity of presentation, the selected submatrices are drawn as continuous blocks (they are not in practice).
        According to a given set of multi-indices $I_k^{(C)}$, the corresponding columns $\my_k^{(C)}$ are computed (shown in green) and transformed by the mapping function~\eqref{eq:ttopt-map} to $\tilde{\my}_k^{(C)}$ (shown in orange), then the rows $I_k^{(R)}$ corresponding to the maximal-volume submatrix $\hat{\my}_k$ (shown in yellow) of $\tilde{\my}_k^{(C)}$ are determined, and then the submatrix $\my_k^{(R)}$ formed by rows $I_k^{(R)}$ (shown in purple) is converted into the next unfolding matrix $\my_{k+1}$.
        The search for the minimum element of the tensor $y_{min}$ is carried out in the maximal-volume submatrix $\hat{\my}_k$.}
    \label{fig:ttopt_scheme_ext}
\end{figure}

The TTOpt algorithm~\cite{sozykin2022ttopt} is based on the modified TT-CAM and the described above property of the maximal-volume submatrix.
It iteratively search for the maximal-volume submatrices in the column and row space of the
implicitly given tensor unfolding matrices.
The illustration for TTOpt algorithm is presented in Figures~\ref{fig:ttopt_scheme} and~\ref{fig:ttopt_scheme_ext}.

The algorithm starts from a given (random) set of $R$ multi-indices $I_1^{(C)}$ for the first unfolding matrix\footnote{
    Precisely, $I_1^{(C)}$ here is a list of $R$ random multi-indices of size $d-1$, which specify positions along modes $k=2$ to $k=d$.
    It may be represented as a matrix $\matr{I}_1^{(C)} \in \set{N}^{R \times (d-1)}$, where each row relates to one selected multi-index $[n_2, n_3, \ldots, n_d]$.
    A similar consideration is also valid for the subsequent index sets chosen for $k$-th ($k = 2, 3, \ldots, d$) unfolding, i.e., $I_k^{(C)} \in \set{N}^{R \times (d-k)}$.

} $\my_1 \in \set{R}^{N_1 \times N_2 \ldots N_d}$ of the implicitly given tensor $\ty \in \set{R}^{N_1 \times N_2 \times \ldots \times N_d}$.
The related submatrix $\my_1^{(C)} \in \set{R}^{N_1 \times R}$ for all positions along the first mode (shown in green in Figures~\ref{fig:ttopt_scheme} and~\ref{fig:ttopt_scheme_ext}) is computed through an explicit call to the target function $\func{f}(\vect{n})$ of discrete argument $\vect{n} = [n_1, n_2, \ldots, n_d]$ being optimized. 
Since maximal-volume submatrices often contain the maximum modulus element, i.e., in absolute value, in~\cite{sozykin2022ttopt} the following dynamic mapping function to find the global minimum is proposed
\begin{equation}\label{eq:ttopt-map}
\func{g}(n)
=
\frac{\pi}{2} -
\func{atan}\left(
    \func{f}(n) - y_{min}
\right),
\end{equation}
where $y_{min}$ is the current best approximation for the minimum element of the tensor (at the beginning of the algorithm, we can, for example, choose $y_{min} = 0$).
According to this mapping function, the columns $\my_1^{(C)}$ are transformed to $\tilde{\my}_1^{(C)}$ (this transformation is shown in orange in Figure~\ref{fig:ttopt_scheme_ext}).

Then, the maxvol algorithm is applied to $\tilde{\my}_1^{(C)}$ to find the maximal-volume submatrix $\hat{\my_1} \in \set{R}^{R \times R}$ (shown in yellow in Figures~\ref{fig:ttopt_scheme} and~\ref{fig:ttopt_scheme_ext}) and the corresponding multi-indices of $R$ rows are stored in the list $I_1^{(R)}$.
If the submatrix $\hat{\my_1}$  contains element less than the current optimum found, then the multi-index $\vect{n}_{min}$ and optimum value $y_{min}$ are updated accordingly.

In case of ordinary CAM for matrices, the next step would be to compute the entire submatrix $\my_1^{(R)}$ corresponding to rows $I_1^{(R)}$ (shown in purple and blue in Figures~\ref{fig:ttopt_scheme} and~\ref{fig:ttopt_scheme_ext}).
But in case of multidimensional tensor, we cannot compute this submatrix, since it contains an exponential number of elements.
The following trick, which was originally proposed in the TT-CAM, is used instead in the TTOpt approach.
The implicit matrix $\my_{1}^{(R)}$ is reshaped as a new matrix $\my_{2} \in \set{R}^{R N_2 \times N_3 \ldots N_d}$, and for this (second unfolding) matrix $\my_{2}$ similar operations are performed.

For a given (random) set of multi-indices $I_2^{(C)}$, the submatrix $\my_2^{(C)} \in \set{R}^{R N_2 \times R}$ of the matrix $\my_{2}$ is computed and transformed according to the formula~\eqref{eq:ttopt-map} to $\tilde{\my}_2^{(C)}$.
Then $R$ rows $I_2^{(R)}$ are selected by maxvol algorithm and related implicit submatrix $\my_{2}^{(R)}$ is similarly reshaped to a new matrix $\my_{3} \in \set{R}^{R N_3 \times N_4 \ldots N_d}$.
The described operations, called sweeps, are continued until the last mode of the initial tensor is reached.
After that, the process is repeated in the opposite direction, sampling now the row indices $I_{k}^{(R)}$ ($k = d, d-1, \ldots, 1$) instead of the column indices $I_{k}^{(C)}$.
These sequences of forward and backward iterations (named sweeps) continue until a budget of $M$ requests to the objective function $\func{f}$ is depleted.

\subsection{Algorithm complexity and tuning parameters}\label{sec:ttopt-complexity}

According to the TTOpt algorithm presented above, for each unfolding matrix $\my_k$ ($k = 1, 2, \ldots, d$), the submatrix $\my_k^{(C)} \in \set{R}^{R N_k \times R}$ is explicitly computed (related submatrices are shown in green in Figures~\ref{fig:ttopt_scheme} and~\ref{fig:ttopt_scheme_ext}).
Within one sweep (forward and backward), $2d$ such matrices are processed.
Hence, $M = 2 \cdot d \cdot \hat{N} \cdot R^2 \cdot T$ requests to the target function $\func{f}$ will be made in total after $T$ sweeps.
Note that the QUBO problem leads to a $d$-dimensional binary tensor\footnote{
    Note that for the binary tensor ($N_1 = N_2 = \ldots = N_d = 2$) and rank $R \geq 2$, all rows in the first unfolding $\my_1$ will be selected, and then the second unfolding $\my_2$ will contain not $R \cdot N_2$ as shown in Figure~\ref{fig:ttopt_scheme}, but $2 \cdot N_2$.
    Similarly, if the rank $R \geq 4$, then all rows will be selected in the second unfolding $\my_2$.
    However, this special case does not actually change the algorithm.
} ($\hat{N} = 2$), hence we have the following relationship between the algorithm parameters
\begin{equation}\label{eq:ttopt_budget}
M = 4 \cdot d \cdot R^2 \cdot T,
\end{equation}
where $M$ is a computational budget, $R$ is a selected rank and $T$ is a number of performed sweeps.

It was shown in~\cite{sozykin2022ttopt} that the computational complexity of the direct operations within the framework of the TTOpt algorithm (excluding requests to the objective function) is $\order{d \cdot \hat{N} \cdot R^3 \cdot T}$, and in case of the QUBO problem it will be
$
\order{d \cdot R^3 \cdot T}
$.
However, if the time of a single call to the objective function $\func{f}$ is significant, then the effort spent on the algorithm's operation will be negligible.
E.g., if one request to the objective function has complexity of order $d^2$ (as in the case of the QUBO problem with a matrix of the shape $d \times d$), then the total complexity related to function evaluations due to~\eqref{eq:ttopt_budget} will be equal to 
$
\order{d^3 \cdot R^2 \cdot T}
$,
and for the case $d^2 \gg R$ this complexity will prevail.

In accordance with the formula~\eqref{eq:ttopt_budget}, the TTOpt algorithm has only two parameters, i.e., $M$ (computational budget) and $R$ (rank), while their choice is limited by the requirement $T \geq 1$ (the algorithm must perform at least one full sweep).
For a given computational budget ($M$), the rank ($R$) should not be too large (algorithm should make at least a few sweeps), on the other hand, the rank should not be too small (this will lead to a decrease in generalizing ability)\footnote{
    In~\cite{sozykin2022ttopt}, for all considered multivariable analytic functions, the rank value equal to $4$ was used.
    For the case $R = 4$ and $T = 1$, the minimum computational budget can be assessed as $M = 64 \cdot d$ in accordance with the formula~\eqref{eq:ttopt_budget}.
    We also note that in the practical implementation of the TTOpt algorithm (see public github repository \url{https://github.com/SkoltechAI/ttopt}), there are a number of other auxiliary parameters, however, as shown by our numerical experiments, they do not significantly affect the result in the QUBO problem.
}.

One of the advantages of TTOpt is the possibility of its application to essentially multidimensional functions and tensors.
In~\cite{sozykin2022ttopt} its performance was checked and compared with powerful evolutionary algorithms on a number of benchmark functions and neural network-optimization problems.
In that work the TTOpt was successfully applied to various 500-dimensional analytical functions, and $2^{25}$ points were selected for each mode as part of discretization. It was followed by quantization of the related $500$-dimensional tensor into a binary tensor of dimensionality $d = 500 \cdot 25 = 12\,500$.
Remarkably, no supercomputing cluster was required for the task. It was reported that \emph{calculations took less than an hour} for any benchmark and were performed \emph{on a regular laptop}.
Thus, potentially the TTOpt approach can be applied to complex QUBO problems with thousands and tens of thousands of features. From the perspective of the feature selection task in recommender systems, it means we can cover many realistic scenarios as the number of available features is unlikely to exceed this threshold.
\section{Results} \label{sec:results}

The source code and instructions can be found in our GitHub repository\footnote{
    See \url{https://github.com/Anonymous-research-entity/RecSys2022-CQFS-TTOpt}.
}.
In the experiments we have used the source code provided by the authors of the original paper\footnote{
    See \url{https://github.com/qcpolimi/CQFS}.
} and only added TTOpt to solve the QUBO problem as a substitution for the quantum annealer. We have followed the results reported in the original paper and ran experiments for the same $\beta$ and $s$ values (see Equations \eqref{eq:ipm} and \eqref{eq:qubo_constraint}) for each of the three datasets: \emph{Xing Challenge 2017} (79 features, 88,984 items and 190,291 users), \emph{The Movies Dataset} (3058 features, 44,711 items and 270,882 users) and \emph{CiteULike-a} (8000 features, 16,980 items and 5551 users). More details on the datasets can be found in the original paper. 

\subsection{Scalability}\label{sec:scalability}
\begin{figure}[h]
    \centering
    \includegraphics[width=0.95\textwidth]{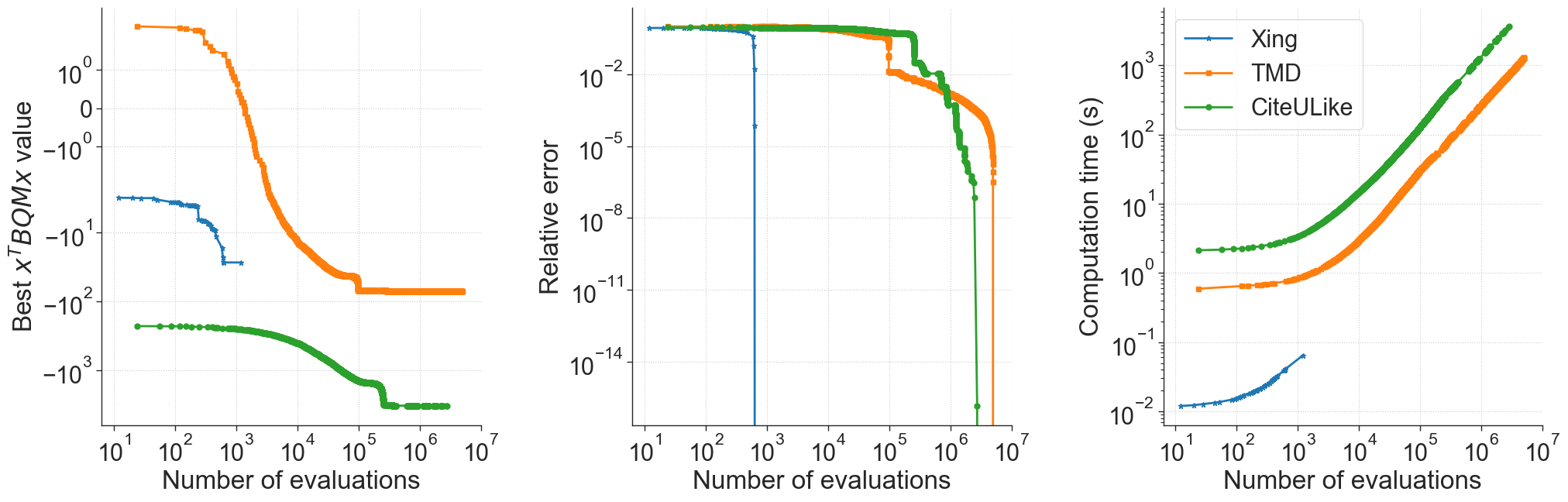}
    \caption{The best found $x^T BQM x$ value (left), relative error from the final optimum after 5 million evaluations (middle) and the computation time (right) depending on the number of evaluations for the three datasets: Xing Challenge 2017 (blue star, 79 features), The Movies Dataset (orange square, 3058 features) and CiteULike-a (green circle, 8000 features). Each point denotes an evaluation when a new optimum is found.}
    \label{fig:ttopt-dataset-performance}
\end{figure}

Since TTOpt algorithm does not include any particular stopping criteria, it is necessary to provide a reasonable evaluation budget for every optimization task. We can pick the rank $R$ and the budget $M$ according to the heuristic from the Equation \eqref{eq:ttopt_budget} in a way that the number of complete sweeps $T$ is at least one. To get better insight on the budget selection and TTOpt performance for the particular datasets, we conducted preliminary experiments on a regular laptop: for each of the datasets we selected values of $\beta$ and $s$ corresponding to the desired fraction of features $p=20\%$ with Item-KNN collaborative model (see Tables \ref{tab:xing_relative}, \ref{tab:tmd_relative}, \ref{tab:cite_relative}) and ran QUBO optimization for the respective $BQM$ matrices (see the equation \eqref{eq:qubo}). Figure \ref{fig:ttopt-dataset-performance} shows the optimum value, relative error from the final optimum (after 5 million evaluations) and computation time depending on the number of evaluations for the three datasets. Each point denotes evaluation when a new optimum is found. The heuristic budget values closely correspond to the regions of a low relative optimum improvement. It can be seen that even without explicit knowledge of the underlying problem (we use TTOpt purely as a black-box optimizer), good solution can be found in a reasonable time. For a small number of features (79), selection is very fast and takes fraction of a second. For a sufficiently large number of features (3058) it takes around 40 seconds. The case with 8000 features takes longer but is still feasible with a run time under 10 minutes. Moreover, in the \cref{sec:results_xing} we compare TTOpt solutions for the particular BQM matrices provided by the authors of the original paper and show that it achieves better optimum value than the one obtained with a quantum annealer.

However, to make sure that the best possible solution is found we increased the computational budgets in our experiments significantly. Table \ref{tab:ttopt-dataset-parameters} describes the values of the TTOpt parameters used for every dataset. In all of the further experiments we utilized a machine with 512 GB RAM and 12 Intel Xeon E5-2670 v3 @ 2.30GHz.

\begin{table}[h]
\begin{tabular}{ll|lll|}
\cline{3-5}
 &  & \multicolumn{3}{c|}{TTOpt} \\ \cline{3-5} 
 &  & \multicolumn{1}{l|}{\multirow{2}{*}{Rank $R$}} & \multicolumn{2}{c|}{Budget $M$} \\ \cline{2-2} \cline{4-5} 
\multicolumn{1}{l|}{} & \# Features & \multicolumn{1}{l|}{} & \multicolumn{1}{l|}{Heuristic} & Used \\ \hline
\multicolumn{1}{|l|}{Xing Challenge 2017} & 79 & \multicolumn{1}{l|}{2} & \multicolumn{1}{l|}{$1.2 \cdot 10^3$} & $10^4$ \\ \hline
\multicolumn{1}{|l|}{The Movies Dataset} & 3058 & \multicolumn{1}{l|}{4} & \multicolumn{1}{l|}{$1.9 \cdot 10^5$} & $10^6$ \\ \hline
\multicolumn{1}{|l|}{CiteULike-a} & 8000 & \multicolumn{1}{l|}{4} & \multicolumn{1}{l|}{$5.1 \cdot 10^5$} & $2 \cdot 10^6$ \\ \hline
\end{tabular}
\caption{Size of the QUBO problem and the values of TTOpt parameters used for each dataset. Details about the TTOpt parameters can be found in the \cref{sec:ttopt-complexity}. \emph{heuristic budget} denotes heuristic value (see the Equation \eqref{eq:ttopt_budget}), whereas \emph{used budget} denotes the over-estimated value that we used in our experiments for certainty.}
\label{tab:ttopt-dataset-parameters}
\end{table}

\subsection{Evaluation}
The originally proposed CQFS feature selection algorithm is designed to operate in a cold-start setting relying purely on the content information from the items. However, it should be trained in a warm-start setting with available collaborative information. To accommodate for this, datasets are sequentially split twice: 1) cold-item split to train, optimize and evaluate feature weighting and selection algorithms with 10\% of interactions for validation and 20\% for testing; then, training and validation data are merged for 2) warm-item split to train and optimize collaborative models that are utilized further by feature selection algorithms with 10\% interactions kept for validation. 

Authors refer to Item-KNN (nearest neighbours), PureSVD (matrix factorization) and $RP^3\beta$ (graph) collaborative models to feed into feature weighting and selection algorithms. Their hyper-parameters are optimized with 50 iterations of Bayesian optimization targeting precision of top-10 recommendations on the warm validation data.

Selected features are subsequently used to build a content-based Item-KNN model using cosine similarity and shrinkage. Hyper-parameters are also optimized with 50 iterations of Bayesian optimization targeting precision of top-10 recommendations on the cold validation data. Once the hyper-parameters are optimized, the final model is built using both training and cold validation data and subsequently evaluated on the cold test data.

CQFS algorithm is compared with the three baseline algorithms: i) Item-KNN with TF-IDF, BM25 or no weighting, ii) Item-KNN selection based on TF-IDF scores, and iii) CFeCBF weighting algorithm incorporating collaborative information. All of the baselines, as well as Item-KNN with CQFS feature selection use Bayesian optimization with 50 iterations for hyper-parameter optimization. More details can be found in the original paper.

We compare reproduced results with the original paper in a relative manner. Train, validation and test splits may differ, thus, we pick the test results of the content-based model as a reference and for each method measure a percentage of improvement in metrics to see if the results exhibit similar behaviour. All of the metrics are evaluated for top-10 recommendations. Additionally, we provide all of the absolute values in supplementary materials.

Table \ref{tab:item_knn_baseline} provides comparison between the original and reproduced baseline results for an Item-KNN content-based model trained using all of the features and evaluated on the test data for all three dataset.

\subsubsection{Xing Challenge 2017}\label{sec:results_xing}
This dataset offers only 79 features but it is very RAM-demanding due to the relatively large number of items of $\sim 89k$ and necessity to construct Item Penalty Matrix from the Equation \eqref{eq:ipm}. We must note that reproduced baseline metrics from the Table \ref{tab:item_knn_baseline} deviate significantly in absolute values from the results reported by the authors of the original paper. 

Table \ref{tab:xing_relative} shows the comparison in changes of metrics for different feature selection/weighting algorithms in percentage relative to the baseline values from the Table \ref{tab:item_knn_baseline}. We can see some discrepancies between the results for TFIDF and CFeCBF algorithms, however, the main concern lies with the CQFS approach. As was mentioned in the \cref{sec:problem}, the soft constraint on the preferred percentage of selected features $p$ (see the Equation \eqref{eq:qubo_constraint}) is not necessarily enforced in practice. In our experiments for 60, 80 and 95\% preferences TTOpt algorithm found optimal, but trivial solutions of all ones, whereas for 40\% preference it selected 97\% of the features.

To obtain more details we contacted the authors of the original paper and they kindly provided us both their selected features and $BQM$ matrices (see Equation \eqref{eq:qubo}) for the reported results. Then, we i) took each of the provided $BQM$ matrices, ii) ran TTOpt to solve the QUBO problem, and iii) compared the obtained optimum values and feature selections with the ones provided by the authors. Parameters for TTOpt were set the same as in the other experiments. The findings are depicted in the Table \ref{tab:xing_bqm_comparison}. We note that for all of the expected amounts of selected features TTOpt found a better solution than the one provided by the authors. Additionally, we can see that the desired amount of selected features is not really attained, and analogously to our main experiments (in all of the cases except $p = 40\%$) selection is trivial and corresponds to all features. Our observations show that trivial selections are caused by a predominance of either negative (small $\beta$) or positive (large $\beta$) values in $BQM$ matrix due to its rather small size. We argue that with a large number of items and a small number of features, penalizing and rewarding components of the Feature Penalty Matrix $FPM$ (see the Equation \eqref{eq:fpm_split}) intersect significantly. Thus, by tuning their weighting $\beta$ we may end up with arbitrary solutions. We do not observe such effects in the other two larger (in terms of features) datasets.

\begin{table}[t]
\begin{tabular}{llllll|}
\cline{3-6}
 & \multicolumn{1}{l|}{} & \multicolumn{1}{l|}{\textbf{NDCG}} & \multicolumn{1}{l|}{\textbf{MAP}} & \multicolumn{1}{l|}{\textbf{Recall}} & \textbf{Coverage} \\ \hline
\multicolumn{1}{|l|}{\multirow{2}{*}{Xing Challenge 2017}} & Reported & 0.0525 & 0.0322 & 0.0680 & 0.9999 \\
\multicolumn{1}{|l|}{} & Reproduced & 0.0840 & 0.0453 & 0.1109 & 1.0000 \\ \hline
\multicolumn{1}{|l|}{\multirow{2}{*}{The Movies Dataset}} & Reported & 0.0856 & 0.0808 & 0.0795 & 0.6637 \\
\multicolumn{1}{|l|}{} & Reproduced & 0.0687 & 0.0649 & 0.0712 & 0.6426 \\ \hline
\multicolumn{1}{|l|}{\multirow{2}{*}{CiteULike-a}} & Reported & 0.2500 & 0.1735 & 0.2803 & 0.9486 \\
\multicolumn{1}{|l|}{} & Reproduced & 0.2467 & 0.1697 & 0.2778 & 0.9444 \\ \hline
\end{tabular}
\caption{Test metrics (@10) for the baseline Item-KNN content-based model evaluated on the three datasets. \emph{Reported} denote values from the original paper, whereas \emph{Reproduced} denote reproduced values.}
\label{tab:item_knn_baseline}
\end{table}

\begin{table}[h]
\small
\begin{tabular}{llccccc|c|c|}
\cline{3-9}
\multicolumn{1}{r}{\textbf{}} & \multicolumn{1}{r|}{\textbf{}} & \multicolumn{1}{r|}{\textbf{Selected, \%}} & \multicolumn{1}{r|}{\textbf{NDCG $\updownarrow$, \%}} & \multicolumn{1}{r|}{\textbf{MAP $\updownarrow$, \%}} & \multicolumn{1}{r|}{\textbf{Recall $\updownarrow$, \%}} & \multicolumn{1}{r|}{\textbf{Coverage $\updownarrow$, \%}} & \textbf{$\beta$} & \textbf{$s$} \\ \hline
\multicolumn{1}{|l|}{\multirow{2}{*}{TFIDF 40\%}} & Reported & - & -94.5 & -95.0 & -93.1 & -75.8 & \multirow{2}{*}{\textbf{-}} & \multirow{2}{*}{\textbf{-}} \\
\multicolumn{1}{|l|}{} & Reproduced & - & -86.9 & -86.8 & -84.6 & -75.9 &  &  \\ \hline
\multicolumn{1}{|l|}{\multirow{2}{*}{TFIDF 60\%}} & Reported & - & -82.5 & -86.0 & -78.4 & -47.5 & \multirow{2}{*}{\textbf{-}} & \multirow{2}{*}{\textbf{-}} \\
\multicolumn{1}{|l|}{} & Reproduced & - & -85.9 & -84.2 & -85.7 & -58.6 &  &  \\ \hline
\multicolumn{1}{|l|}{\multirow{2}{*}{TFIDF 80\%}} & Reported & - & -52.4 & -61.8 & -46.9 & -10.2 & \multirow{2}{*}{\textbf{-}} & \multirow{2}{*}{\textbf{-}} \\
\multicolumn{1}{|l|}{} & Reproduced & - & -51.0 & -43.0 & -55.5 & -10.6 &  &  \\ \hline
\multicolumn{1}{|l|}{\multirow{2}{*}{TFIDF 95\%}} & Reported & - & -6.1 & -10.9 & -0.6 & -0.1 & \multirow{2}{*}{\textbf{-}} & \multirow{2}{*}{\textbf{-}} \\
\multicolumn{1}{|l|}{} & Reproduced & - & -0.2 & 4.1 & 1.5 & 0.0 &  &  \\ \hline
\multicolumn{1}{|l|}{\multirow{2}{*}{CFeCBF ItemKNN}} & Reported & - & -38.3 & -46.6 & -28.1 & -1.3 & \multirow{2}{*}{\textbf{-}} & \multirow{2}{*}{\textbf{-}} \\
\multicolumn{1}{|l|}{} & Reproduced & - & -31.6 & -20.6 & -29.4 & -0.1 &  &  \\ \hline
\multicolumn{1}{|l|}{\multirow{2}{*}{CFeCBF PureSVD}} & Reported & - & -96.6 & -97.2 & -95.6 & -27.3 & \multirow{2}{*}{\textbf{-}} & \multirow{2}{*}{\textbf{-}} \\
\multicolumn{1}{|l|}{} & Reproduced & - & -94.1 & -94.4 & -93.5 & -63.8 &  &  \\ \hline
\multicolumn{1}{|l|}{\multirow{2}{*}{CFeCBF RP$^3\beta$}} & Reported & - & -14.1 & -9.9 & -12.4 & -0.1 & \multirow{2}{*}{\textbf{-}} & \multirow{2}{*}{\textbf{-}} \\
\multicolumn{1}{|l|}{} & Reproduced & - & -15.7 & -4.3 & -16.5 & -0.5 &  &  \\ \hline
\multicolumn{1}{|l|}{\multirow{2}{*}{CQFS ItemKNN 40\%}} & Reported & 85 & -13.5 & -14.3 & -14.7 & -0.4 & \multirow{2}{*}{0.001} & \multirow{2}{*}{1000} \\
\multicolumn{1}{|l|}{} & Reproduced & 97 & -11.5 & -12.7 & -9.1 & 0.0 &  &  \\ \hline
\multicolumn{1}{|l|}{\multirow{2}{*}{CQFS ItemKNN 60\%}} & Reported & 82 & -2.3 & -4.7 & -0.3 & 0.0 & \multirow{2}{*}{0.0001} & \multirow{2}{*}{10} \\
\multicolumn{1}{|l|}{} & Reproduced & 100 & 1.5 & 1.2 & 1.4 & 0.0 &  &  \\ \hline
\multicolumn{1}{|l|}{\multirow{2}{*}{CQFS ItemKNN 80\%}} & Reported & 84 & 4.0 & -2.5 & 9.4 & 0.0 & \multirow{2}{*}{0.001} & \multirow{2}{*}{100} \\
\multicolumn{1}{|l|}{} & Reproduced & 100 & 1.5 & 1.2 & 1.4 & 0.0 &  &  \\ \hline
\multicolumn{1}{|l|}{\multirow{2}{*}{CQFS ItemKNN 95\%}} & Reported & 89 & -1.3 & -2.5 & -1.3 & -0.3 & \multirow{2}{*}{0.0001} & \multirow{2}{*}{1000} \\
\multicolumn{1}{|l|}{} & Reproduced & 100 & 1.5 & 1.2 & 1.4 & 0.0 &  &  \\ \hline
\end{tabular}
\caption{Changes of test metrics (@10) in percentage relative to the the baseline Item-KNN content-based model from the Table \ref{tab:item_knn_baseline} for several feature selection/weighting algorithms evaluated on the Xing Challenge 2017 dataset. \emph{Reported} denote values from the original paper, whereas \emph{Reproduced} denote values obtained by replacing a quantum annealer with TTOpt algorithm to solve the QUBO problem (see the Equation \eqref{eq:qubo}). \emph{Selected} shows how many features were indeed picked, where the values for the original paper were kindly provided by the authors upon our request.}
\label{tab:xing_relative}
\end{table}

\begin{table}[h]
\begin{tabular}{ll|l|l|}
\cline{3-4}
& \textbf{}  & Selected, \% & $x^T BQM x$        \\ \hline
\multicolumn{1}{|c|}{\multirow{2}{*}{\textbf{$\beta=10^{-3},\,s=10^3,\,p=40 \%$}}} & Reported   & 85           & -26.4616 \\
\multicolumn{1}{|c|}{}                                                             & Reproduced & 97           & \textbf{-27.0554} \\ \hline
\multicolumn{1}{|l|}{\multirow{2}{*}{$\beta=10^{-4},\,s=10,\,p=60 \%$}}            & Reported   & 82           & -26.7404          \\
\multicolumn{1}{|l|}{}                                                             & Reproduced & 100          & \textbf{-27.4633} \\ \hline
\multicolumn{1}{|l|}{\multirow{2}{*}{$\beta=10^{-3},\,s=10^2,\,p=80 \%$}}          & Reported   & 84           & -26.6192          \\
\multicolumn{1}{|l|}{}                                                             & Reproduced & 100          & \textbf{-27.1378} \\ \hline
\multicolumn{1}{|l|}{\multirow{2}{*}{$\beta=10^{-4},\,s=10^3,\,p=95 \%$}}          & Reported   & 89           & -27.3326          \\
\multicolumn{1}{|l|}{}                                                             & Reproduced & 100          & \textbf{-27.6798} \\ \hline
\end{tabular}
\caption{Comparison of the percentage of selected features and respective values of the quadratic forms for the optimal parameters reported in the original paper for the Xing Challenge 2017 dataset. $BQM$ matrices and feature selections were kindly provided by the authors. \emph{Reported} denote original results, whereas \emph{Reproduced} denote results obtained from the provided BQM matrices with TTOpt optimization.}
\label{tab:xing_bqm_comparison}
\end{table}

\subsubsection{The Movies Dataset}\label{sec:results_tmd}
This dataset contains larger number of features of $\sim 3k$ and is less RAM demanding due to a smaller number of items $\sim 45k$. Similarly to the Xing Challenge 2017 dataset we can note significant discrepancies in absolute values between the reported and reproduced metrics from the Table \ref{tab:item_knn_baseline} for the baseline model.

Table \ref{tab:tmd_relative} shows the comparison in changes of metrics for different feature selection/weighting algorithms in percentage relative to the baseline values from the Table \ref{tab:item_knn_baseline}. We can see some discrepancies between the results for TFIDF and CFeCBF algorithms, as well as CQFS based on Item-KNN collaborative model for $p=40\%, 80\%, 95\%$ and CQFS based on $RP^3\beta$ collaborative model for $p=20\%, 30\%, 40\%$. Other results seems to be consistent. Feature selection with CQFS based on PureSVD collaborative model yielded the best results in both reported and reproduced cases, improving on a thin margin over the baseline with 43\% features selected. Analogously to the Xing Challenge 2017 dataset, the desired fraction of selected features $p$ is not necessarily attained in practice. However, we can see that CQFS based on an Item-KNN collaborative model coincides almost perfectly (except for $p=20\%$) with the number of selected features. We argue that in this case penalty associated with a difference in the number of selected and desired features from the Equation \ref{eq:qubo_constraint} is predominant over the feature penalization term itself.

\begin{table}[h]
\begin{tabular}{llccccc|c|c|}
\cline{3-9}
\multicolumn{1}{r}{\textbf{}} & \multicolumn{1}{c|}{\textbf{}} & \multicolumn{1}{c|}{\textbf{Selected, \%}} & \multicolumn{1}{c|}{\textbf{NDCG $\updownarrow$, \%}} & \multicolumn{1}{c|}{\textbf{MAP $\updownarrow$, \%}} & \multicolumn{1}{c|}{\textbf{Recall $\updownarrow$, \%}} & \textbf{Coverage $\updownarrow$, \%} & \textbf{$\beta$} & \textbf{$s$} \\ \hline
\multicolumn{1}{|l|}{\multirow{2}{*}{TFIDF 40\%}} & Reported & - & -68.6 & -58.7 & -58.3 & -12.7 & \multirow{2}{*}{-} & \multirow{2}{*}{-} \\
\multicolumn{1}{|l|}{} & Reproduced & - & -66.4 & -52.5 & -60.6 & -12.1 &  &  \\ \hline
\multicolumn{1}{|l|}{\multirow{2}{*}{TFIDF 60\%}} & Reported & - & -30.5 & -34.9 & -33.0 & -6.7 & \multirow{2}{*}{-} & \multirow{2}{*}{-} \\
\multicolumn{1}{|l|}{} & Reproduced & - & -53.4 & -44.3 & -44.8 & -6.5 &  &  \\ \hline
\multicolumn{1}{|l|}{\multirow{2}{*}{TFIDF 80\%}} & Reported & - & -26.1 & -31.2 & -30.1 & -1.6 & \multirow{2}{*}{-} & \multirow{2}{*}{-} \\
\multicolumn{1}{|l|}{} & Reproduced & - & -24.4 & -23.7 & -20.9 & -0.9 &  &  \\ \hline
\multicolumn{1}{|l|}{\multirow{2}{*}{TFIDF 95\%}} & Reported & - & -22.3 & -25.0 & -21.5 & -1.4 & \multirow{2}{*}{-} & \multirow{2}{*}{-} \\
\multicolumn{1}{|l|}{} & Reproduced & - & -26.8 & -30.1 & -29.7 & -0.7 &  &  \\ \hline
\multicolumn{1}{|l|}{\multirow{2}{*}{CFeCBF ItemKNN}} & Reported &  & -19.2 & -26.7 & -26.0 & -2.3 & \multirow{2}{*}{-} & \multirow{2}{*}{-} \\
\multicolumn{1}{|l|}{} & Reproduced & - & -36.8 & -34.7 & -42.5 & -2.3 &  &  \\ \hline
\multicolumn{1}{|l|}{\multirow{2}{*}{CFeCBF PureSVD}} & Reported & - & -20.4 & -32.8 & -28.6 & -4.5 & \multirow{2}{*}{-} & \multirow{2}{*}{-} \\
\multicolumn{1}{|l|}{} & Reproduced & - & -63.0 & -61.8 & -71.6 & -16.1 &  &  \\ \hline
\multicolumn{1}{|l|}{\multirow{2}{*}{CFeCBF RP$^3\beta$}} & Reported & - & -34.2 & -38.3 & -38.2 & -1.4 & \multirow{2}{*}{-} & \multirow{2}{*}{-} \\
\multicolumn{1}{|l|}{} & Reproduced & - & -44.5 & -37.6 & -46.6 & -1.3 &  &  \\ \hline
\multicolumn{1}{|l|}{\multirow{2}{*}{CQFS ItemKNN 20\%}} & Reported & 26 & -7.9 & -7.7 & -10.3 & -0.2 & \multirow{2}{*}{0.0001} & \multirow{2}{*}{100} \\
\multicolumn{1}{|l|}{} & Reproduced & 22 & -7.9 & -5.8 & -7.8 & -0.4 &  &  \\ \hline
\multicolumn{1}{|l|}{\multirow{2}{*}{CQFS ItemKNN 30\%}} & Reported & 30 & -4.3 & -4.5 & -5.1 & -0.1 & \multirow{2}{*}{0.0001} & \multirow{2}{*}{1000} \\
\multicolumn{1}{|l|}{} & Reproduced & 30 & -3.2 & -5.5 & -4.9 & -0.3 &  &  \\ \hline
\multicolumn{1}{|l|}{\multirow{2}{*}{CQFS ItemKNN 40\%}} & Reported & 43 & -0.4 & -0.8 & -0.2 & -0.1 & \multirow{2}{*}{0.001} & \multirow{2}{*}{100} \\
\multicolumn{1}{|l|}{} & Reproduced & 40 & -6.9 & -14.5 & -6.7 & -1.1 &  &  \\ \hline
\multicolumn{1}{|l|}{\multirow{2}{*}{CQFS ItemKNN 60\%}} & Reported & 62 & 1.3 & 1.2 & 2.5 & 0.1 & \multirow{2}{*}{0.0001} & \multirow{2}{*}{100} \\
\multicolumn{1}{|l|}{} & Reproduced & 60 & -5.3 & -3.7 & -4.4 & -0.3 &  &  \\ \hline
\multicolumn{1}{|l|}{\multirow{2}{*}{CQFS ItemKNN 80\%}} & Reported & 81 & 1.1 & 1.1 & 2.0 & 0.1 & \multirow{2}{*}{0.001} & \multirow{2}{*}{100} \\
\multicolumn{1}{|l|}{} & Reproduced & 80 & -10.6 & -17.4 & -10.6 & -1.0 &  &  \\ \hline
\multicolumn{1}{|l|}{\multirow{2}{*}{CQFS ItemKNN 95\%}} & Reported & 95 & 0.6 & 0.6 & 0.7 & 0.0 & \multirow{2}{*}{0.001} & \multirow{2}{*}{1000} \\
\multicolumn{1}{|l|}{} & Reproduced & 95 & -9.3 & -16.6 & -3.8 & -0.8 &  &  \\ \hline
\multicolumn{1}{|l|}{\multirow{2}{*}{CQFS PureSVD 20\%}} & Reported & 34 & -2.1 & -2.2 & -2.6 & -0.1 & \multirow{2}{*}{0.0001} & \multirow{2}{*}{100} \\
\multicolumn{1}{|l|}{} & Reproduced & 47 & 0.3 & 0.0 & 0.1 & -0.1 &  &  \\ \hline
\multicolumn{1}{|l|}{\multirow{2}{*}{CQFS PureSVD 30\%}} & Reported & 40 & 0.4 & 0.4 & 0.9 & -0.0 & \multirow{2}{*}{0.0001} & \multirow{2}{*}{100} \\
\multicolumn{1}{|l|}{} & Reproduced & 52 & 0.3 & 0.1 & 0.1 & -0.0 &  &  \\ \hline
\multicolumn{1}{|l|}{\multirow{2}{*}{CQFS PureSVD 40\%}} & Reported & 41 & 0.1 & 0.1 & 0.4 & -0.0 & \multirow{2}{*}{0.0001} & \multirow{2}{*}{1000} \\
\multicolumn{1}{|l|}{} & Reproduced & 43 & 0.8 & 0.2 & 0.5 & -0.1 &  &  \\ \hline
\multicolumn{1}{|l|}{\multirow{2}{*}{CQFS PureSVD 60\%}} & Reported & 60 & 0.6 & 1.1 & 1.5 & 0.1 & \multirow{2}{*}{0.0001} & \multirow{2}{*}{1000} \\
\multicolumn{1}{|l|}{} & Reproduced & 61 & 1.0 & 0.5 & 0.9 & -0.1 &  &  \\ \hline
\multicolumn{1}{|l|}{\multirow{2}{*}{CQFS PureSVD 80\%}} & Reported & 80 & -1.8 & -1.6 & -2.4 & 0.0 & \multirow{2}{*}{0.001} & \multirow{2}{*}{1000} \\
\multicolumn{1}{|l|}{} & Reproduced & 80 & 0.1 & -0.0 & -0.0 & -0.0 &  &  \\ \hline
\multicolumn{1}{|l|}{\multirow{2}{*}{CQFS PureSVD 95\%}} & Reported & 96 & 0.8 & 0.8 & 1.2 & 0.1 & \multirow{2}{*}{0.001} & \multirow{2}{*}{100} \\
\multicolumn{1}{|l|}{} & Reproduced & 99 & -0.1 & -0.1 & -0.2 & 0.0 &  &  \\ \hline
\multicolumn{1}{|l|}{\multirow{2}{*}{CQFS RP$^3\beta$ 20\%}} & Reported & 24 & -18.0 & -10.9 & -16.2 & 0.3 & \multirow{2}{*}{0.001} & \multirow{2}{*}{100} \\
\multicolumn{1}{|l|}{} & Reproduced & 35 & -8.6 & -6.3 & -8.3 & -0.3 &  &  \\ \hline
\multicolumn{1}{|l|}{\multirow{2}{*}{CQFS RP$^3\beta$ 30\%}} & Reported & 33 & -17.1 & -9.9 & -14.7 & 0.4 & \multirow{2}{*}{0.0001} & \multirow{2}{*}{100} \\
\multicolumn{1}{|l|}{} & Reproduced & 43 & -5.4 & -4.0 & -5.1 & -0.2 &  &  \\ \hline
\multicolumn{1}{|l|}{\multirow{2}{*}{CQFS RP$^3\beta$ 40\%}} & Reported & 40 & -14.7 & -8.8 & -12.6 & 0.4 & \multirow{2}{*}{0.0001} & \multirow{2}{*}{1000} \\
\multicolumn{1}{|l|}{} & Reproduced & 42 & -7.3 & -5.6 & -6.9 & -0.3 &  &  \\ \hline
\multicolumn{1}{|l|}{\multirow{2}{*}{CQFS RP$^3\beta$ 60\%}} & Reported & 62 & -0.9 & -1.6 & -0.6 & 0.2 & \multirow{2}{*}{0.0001} & \multirow{2}{*}{100} \\
\multicolumn{1}{|l|}{} & Reproduced & 67 & -1.5 & -1.4 & -1.5 & -0.1 &  &  \\ \hline
\multicolumn{1}{|l|}{\multirow{2}{*}{CQFS RP$^3\beta$ 80\%}} & Reported & 80 & -1.4 & -2.1 & -1.6 & 0.2 & \multirow{2}{*}{0.0001} & \multirow{2}{*}{1000} \\
\multicolumn{1}{|l|}{} & Reproduced & 80 & -2.0 & -1.3 & -1.4 & -0.1 &  &  \\ \hline
\multicolumn{1}{|l|}{\multirow{2}{*}{CQFS RP$^3\beta$ 95\%}} & Reported & 95 & -1.6 & -1.5 & -2.2 & 0.0 & \multirow{2}{*}{0.001} & \multirow{2}{*}{100} \\
\multicolumn{1}{|l|}{} & Reproduced & 98 & -0.1 & -0.1 & -0.2 & 0.0 &  &  \\ \hline
\end{tabular}
\caption{Changes of test metrics (@10) in percentage relative to the the baseline Item-KNN content-based model (see Table \ref{tab:item_knn_baseline}) for several feature selection/weighting algorithms evaluated on The Movies Dataset. \emph{Reported} denote values from the original paper, whereas \emph{Reproduced} denote values obtained by replacing a quantum annealer with TTOpt algorithm to solve the QUBO problem (see the Equation \eqref{eq:qubo}). \emph{Selected} shows how many features were indeed picked, where the values for the original paper were kindly provided by the authors upon our request.}
\label{tab:tmd_relative}
\end{table}

\subsubsection{CiteULike-a}\label{sec:results_cite}
This dataset contains even larger number of features of $8k$ and smaller number of items $\sim 17k$. This dataset shows the most consistency between the reported and reproduced results from the Table \ref{tab:item_knn_baseline} for the baseline model.

Table \ref{tab:cite_relative} shows the comparison in changes of metrics for different feature selection/weighting algorithms in percentage relative to the baseline values from the Table \ref{tab:item_knn_baseline}. We can see some only small discrepancies between the results for CQFS with the desired fraction of selected features $p=60\%, 80\%, 95\%$. Decreasing the number of selected features for this dataset leads to monotonic deacrese in recommendations quality. 

\begin{table}[h]
\begin{tabular}{llccccc|c|c|}
\cline{3-9}
\multicolumn{1}{r}{\textbf{}} & \multicolumn{1}{r|}{\textbf{}} & \multicolumn{1}{r|}{\textbf{Selected, \%}} & \multicolumn{1}{r|}{\textbf{NDCG $\updownarrow$, \%}} & \multicolumn{1}{r|}{\textbf{MAP $\updownarrow$, \%}} & \multicolumn{1}{r|}{\textbf{Recall $\updownarrow$, \%}} & \multicolumn{1}{r|}{\textbf{Coverage $\updownarrow$, \%}} & \textbf{$\beta$} & \textbf{$s$} \\ \hline
\multicolumn{1}{|l|}{\multirow{2}{*}{TFIDF 40\%}} & Reported & - & -45.2 & -48.6 & -48.5 & -2.9 & \multirow{2}{*}{-} & \multirow{2}{*}{-} \\
\multicolumn{1}{|l|}{} & Reproduced & - & -46.1 & -49.6 & -49.3 & -8.3 &  &  \\ \hline
\multicolumn{1}{|l|}{\multirow{2}{*}{TFIDF 60\%}} & Reported & - & -30.3 & -33.0 & -31.8 & 1.0 & \multirow{2}{*}{-} & \multirow{2}{*}{-} \\
\multicolumn{1}{|l|}{} & Reproduced & - & -31.4 & -35.1 & -33.2 & -4.2 &  &  \\ \hline
\multicolumn{1}{|l|}{\multirow{2}{*}{TFIDF 80\%}} & Reported & - & -18.1 & -20.7 & -18.3 & -1.6 & \multirow{2}{*}{-} & \multirow{2}{*}{-} \\
\multicolumn{1}{|l|}{} & Reproduced & - & -19.7 & -23.2 & -20.4 & -0.5 &  &  \\ \hline
\multicolumn{1}{|l|}{\multirow{2}{*}{TFIDF 95\%}} & Reported & - & -7.6 & -7.8 & -9.2 & 1.9 & \multirow{2}{*}{-} & \multirow{2}{*}{-} \\
\multicolumn{1}{|l|}{} & Reproduced & - & -5.5 & -6.6 & -6.2 & -1.0 &  &  \\ \hline
\multicolumn{1}{|l|}{\multirow{2}{*}{CFeCBF ItemKNN}} & Reported & - & -7.2 & -8.6 & -6.8 & -2.7 & \multirow{2}{*}{-} & \multirow{2}{*}{-} \\
\multicolumn{1}{|l|}{} & Reproduced & - & -6.5 & -8.8 & -6.3 & 0.0 &  &  \\ \hline
\multicolumn{1}{|l|}{\multirow{2}{*}{CFeCBF PureSVD}} & Reported & - & -17.4 & -20.8 & -17.1 & -5.4 & \multirow{2}{*}{-} & \multirow{2}{*}{-} \\
\multicolumn{1}{|l|}{} & Reproduced & - & -16.3 & -19.6 & -15.1 & -5.3 &  &  \\ \hline
\multicolumn{1}{|l|}{\multirow{2}{*}{CFeCBF RP$^3\beta$}} & Reported & - & -11.8 & -14.2 & -11.7 & -4.2 & \multirow{2}{*}{-} & \multirow{2}{*}{-} \\
\multicolumn{1}{|l|}{} & Reproduced & - & -13.4 & -16.3 & -12.5 & -3.3 &  &  \\ \hline
\multicolumn{1}{|l|}{\multirow{2}{*}{CQFS ItemKNN 20\%}} & Reported & 33 & -10.3 & -11.9 & -10.5 & 1.9 & \multirow{2}{*}{0.0001} & \multirow{2}{*}{100} \\
\multicolumn{1}{|l|}{} & Reproduced & 33 & -13.3 & -15.7 & -13.8 & -1.3 &  &  \\ \hline
\multicolumn{1}{|l|}{\multirow{2}{*}{CQFS ItemKNN 30\%}} & Reported & 40 & -9.1 & -10.7 & -8.6 & 0.6 & \multirow{2}{*}{0.0001} & \multirow{2}{*}{100} \\
\multicolumn{1}{|l|}{} & Reproduced & 39 & -10.4 & -12.1 & -10.8 & -1.0 &  &  \\ \hline
\multicolumn{1}{|l|}{\multirow{2}{*}{CQFS ItemKNN 40\%}} & Reported & 48 & -8.2 & -9.5 & -8.3 & 0.3 & \multirow{2}{*}{0.0001} & \multirow{2}{*}{100} \\
\multicolumn{1}{|l|}{} & Reproduced & 45 & -9.5 & -10.6 & -10.3 & -1.4 &  &  \\ \hline
\multicolumn{1}{|l|}{\multirow{2}{*}{CQFS ItemKNN 60\%}} & Reported & 65 & -3.2 & -4.0 & -2.8 & 0.4 & \multirow{2}{*}{0.0001} & \multirow{2}{*}{100} \\
\multicolumn{1}{|l|}{} & Reproduced & 63 & -7.7 & -8.5 & -8.7 & -0.5 &  &  \\ \hline
\multicolumn{1}{|l|}{\multirow{2}{*}{CQFS ItemKNN 80\%}} & Reported & 82 & -3.1 & -4.2 & -3.0 & 0.5 & \multirow{2}{*}{0.01} & \multirow{2}{*}{100} \\
\multicolumn{1}{|l|}{} & Reproduced & 81 & -6.6 & -7.7 & -7.6 & -0.5 &  &  \\ \hline
\multicolumn{1}{|l|}{\multirow{2}{*}{CQFS ItemKNN 95\%}} & Reported & 95 & 0.0 & -0.2 & 0.5 & 0.9 & \multirow{2}{*}{0.001} & \multirow{2}{*}{1000} \\
\multicolumn{1}{|l|}{} & Reproduced & 95 & -3.8 & -4.8 & -4.3 & 0.3 &  &  \\ \hline
\end{tabular}
\caption{Changes of test metrics (@10) in percentage relative to the baseline Item-KNN content-based model (see Table \ref{tab:item_knn_baseline}) for several feature selection/weighting algorithms evaluated on the CiteULike-a dataset. \emph{Reported} denote values from the original paper, whereas \emph{Reproduced} denote values obtained by replacing a quantum annealer with TTOpt algorithm to solve the QUBO problem (see the Equation \eqref{eq:qubo}). \emph{Selected} shows how many features were indeed picked, where the values for the original paper were kindly provided by the authors upon our request.}
\label{tab:cite_relative}
\end{table}

\section{Conclusion}
We have revisited a recently proposed optimization problem that tackles the cold-start recommendation problem.
In particular, we reproduced the target QUBO optimization function proposed in~\cite{nembrini2021feature} and measured its performance in the same downstream tasks and datasets considered in the original paper.
However, we used a different optimizer, namely a tensor network-inspired method TTOpt~\cite{sozykin2022ttopt} that runs on a classical computer.
We show that such a classical black-box optimization method can run in reasonable time, even for large problems with thousands of unknowns, while achieving competitive results.
All in all, our study provides a new angle on the attractiveness of state-of-the-art quantum-based solutions for recommender systems vs. its classical counterparts.

\clearpage
\bibliographystyle{ACM-Reference-Format}
\bibliography{acmart}

\end{document}